%% file: cond.tex
\documentclass[aps,prd,showpacs,superscriptaddress,reprint]{revtex4-1}
\usepackage{amsmath,amsfonts,amssymb}
\usepackage{dsfont,graphicx,bm}
\graphicspath{{Plots/}}%{../}}
\newcommand{\TE}{\textsf{TE}}
\newcommand{\TM}{\textsf{TM}}
\newcommand{\CP}{\textsf{CP}}

\newcommand{\te}{\textsf{te}}
\newcommand{\tm}{\textsf{tm}}

\newcommand{\gs}{\textsl{g}}
\newcommand{\bs}{\begin{subequations}}
\newcommand{\es}{\end{subequations}}
\newcommand{\adb}{\allowdisplaybreaks } 
\newcommand{\ann}{\adb \nonumber \\}

\newcommand{\kb}{\bm{k}}
\newcommand{\vb}{\bm{v}}
\newcommand{\ii}{\mathrm{i}}
\usepackage{color} \definecolor{darkgreen}{rgb}{0,.5,0}
\usepackage[colorlinks,filecolor=blue,citecolor=darkgreen,unicode]{hyperref}
\begin{document}
%--------------------------------------------------------------------
\title{The low temperature behavior the Casimir-Polder energy for conductive plane}
\author{Nail Khusnutdinov}\email{nail.khusnutdinov@gmail.com}
\affiliation{Centro de Matem\'atica, Computa\c{c}\~ao e Cogni\c{c}\~ao, Universidade Federal do ABC, 09210-170 Santo Andr\'e, SP, Brazil}
\affiliation{Institute of Physics, Kazan Federal University, Kremlevskaya 18, Kazan, 420008, Russia}
\author{Natalia Emelianova}\email{natalia.emelianova@ufabc.edu.br}
\affiliation{Centro de Matem\'atica, Computa\c{c}\~ao e Cogni\c{c}\~ao, Universidade Federal do ABC, 09210-170 Santo Andr\'e, SP, Brazil}
\date{\today}
\begin{abstract}
The low temperature expansion of the free energy of atom/plane system is considered for general symmetric form of tensor conductivity of the plane. It is shown that the first correction is proportional to second order of the temperature $\sim T^2$ and comes from \TM\ mode.  The agreement of the expansion and exact expressions for different models of conductivity is numerically demonstrated.   
\end{abstract}  
\maketitle 

\section{Introduction}

Van der Waals dispersion forces play an important role in different physical, biological as well as chemical phenomena \cite{Bordag:2009:ACE,*Parsegian:2006:VdWFHBCEP,*Milonni:1994:QVItQE,*Woods:2016:MpoCavdWi}. In the case of interaction between particle and plate it is commonly referred to as the Casimir-Polder force \cite{Casimir:1948:TIrLdWf}. The van der Waals force is very important for interaction of graphene with microparticles \cite{Bondarev:2005:vdWcadcn,*Blagov:2007:vdWibmscn,*Churkin:2010:CohaDmodibgaHHoNa,%
*Khusnutdinov:2011:vdWibaaaasps,*Khusnutdinov:2012:tcpiawsps,*Ribeiro:2013:Svfwg,*Judd:2011:qruaftfgsh}, where finite conductivity of graphene plays essential role \cite{Khusnutdinov:2016:CPefasocp,*Kashapov:2016:TCefpls}. At a short range the
energy rises as the third power of inverse distance between the microparticle and the plate. The retardation of the interaction should be taken into account at large distances and the interaction energy falls down as the fourth power of distance. At separations larger than a few micrometers, thermal effects become dominant. 

Thermal corrections for van der Waalse energy of the system atom/slab and atom/graphene were considered in Ref. \cite{Bezerra:2008:Ltaiatqr,*Chaichian:2012:TCiodawg,*Bordag:2014:lteLf,*Khusnutdinov:2018:tccpinp2ddm}. It was shown that  the correction to the Casimir-Polder free energy is proportional to forth degree of temperature $\sim T^4$ in the case atom and ideal plane. Different models were considered to describe graphene namely, i) hydrodynamical model \cite{Barton:2004:Cesps,*Barton:2005:CefpsIE}, ii) density-density correlation function \cite{Klimchitskaya:2014:tadCigddcfvpt} and iii) the Dirac model \cite{CastroNeto:2009:epg}. In framework of Dirac model was found tensor of conductivity of graphene with and temporal and spatial dispersions and dependence of temperature and chemical potential \cite{Bordag:2009:CibapcagdbtDm,*Fialkovsky:2011:FCefg,*Bordag:2016:ECefdg,*Bordag:2017:EECefdg}. 

In the present paper we consider low temperature expansion of the free Casimir-Polder energy for atom/plane system taking into account general symmetric form of the plane's conductivity tensor. We found that the first correction is quadratic over temperature $\sim T^2$. Numerically we justify low temperature expansion for three different models of conductivity -- constant conductivity model, the Drude-Lorentz model and conductivity calculated in the context of polarization tensor approach. 

The paper is organized as follows. In Sec. \ref{Sec:Structure}, we briefly consider general structure of the conductivity tensor. Section \ref{Sec:CP} presents different representations of the Casimir-Polder energy. In Sec. \ref{Sec:LowT} we derive the main expressions for low temperature expansion and in Sec. \ref{Sec:Numeric} we numerically compare exact expressions and low temperature approximations.  In Sec. \ref{Sec:Conclusion} we discuss the obtained results. Appendix \ref{Sec:Models} devotes for different models of graphene's conductivity and in Appendix \ref{Sec:AppB} we obtain low temperature expansion of the conductivity. 

\section{The structure of the tensor conductivity} \label{Sec:Structure}

Let us consider conductive $2D$ infinitely thin layer positioned perpendicular to axes $z$. We suppose that the anisotropic Ohm low, $\mathbf{j}_s = \bm{\sigma} \mathbf{E}$, is satisfied on the plane, where $\mathbf{j}_s$ is surface current and $\bm{\sigma}$ is conductivity tensor. The latter, in general, depends on the frequency $\omega$, wave vector $\kb$, velocity $\vb$ and other scalar parameters such as temperature $T$ and chemical potential $\mu$. It has the following structure \cite{Zeitlin:1995:QnfdqHe} ($i,j=x,y$) 
\begin{equation}
\sigma_{ij} =  A \delta_{ij}  + B k_ik_j + C k_{(i} v_{j)} + D v_i v_j + E \varepsilon_{ij}, \label{eq:sigmaT}
\end{equation}
where the constant $E$ describes parity-odd part of conductivity \cite{Fialkovsky:2009:poeaprig,*Fialkovsky:2012:QFTiG} and $\varepsilon_{ij}$ is complete antisymmetric tensor. We consider here parity-even part of conductivity without velocity $\vb=0$: 
\begin{equation}
\sigma_{ij} =  A \delta_{ij}  + B k_ik_j. \label{eq:sigmaT}
\end{equation}  

The eigenvalues of this tensor,
\begin{equation}
\sigma^\te = A,\ \sigma^\tm = A + k^2 B, \label{eq:eigen}
\end{equation}
are the conductivities of \TE\ and \TM\ modes. Indeed, boundary conditions for \TE\ ($E_z =0 $) and \TM\ ($H_z =0$) modes have the following form
\begin{eqnarray}
\TE&:& [H_z] =0,\ [H'_z] = +4\pi \ii \omega \sigma^\te H_z, \ann
\TM&:& [E'_z] =0,\ [E_z]  = -\frac{4\pi \ii }{\omega}  \sigma^\tm E'_z, \label{eq:BL}
\end{eqnarray}
where $[f]= f_{z-0} - f_{z+0}$ means jump of function at the layer. Therefore, we observe that eigenvalues $\sigma^\te$ and $\sigma^\tm$ play the role of conductivity for \TE\ and \TM\ modes, respectively. 

Using boundary conditions \eqref{eq:BL} for scattering process we obtain the transmission and reflection coefficients 
\begin{gather}
r^\te (\omega,k_z) = -\frac{\eta^\te}{\eta^\te + \frac{k_z}{\omega}},\ t^\te = 1 + r^\te,\ann
r^\tm (\omega,k_z) = \frac{\eta^\tm}{\eta^\tm + \frac{\omega}{k_z}},\ t^\tm  = 1 - r^\tm, \label{eq:transmission}
\end{gather}
where $k_z = \sqrt{\omega^2 - k_\perp^2}$ and $\eta^{\te,\tm} = 2\pi \sigma^{\te,\tm}$.

\section{The Casimir-Polder free energy} \label{Sec:CP}

The system under consideration consists of atom and conductive plane with distance $a$ between atom and plane. Using the rarefied procedure of Lifshitz \cite{Lifshitz:1956:tmafbs} the Casimir-Polder ($\CP$) energy can be given as a sum of \TM\ and \TE\ contributions \cite{Khusnutdinov:2016:CPefasocp,*Kashapov:2016:TCefpls},
\begin{eqnarray}
\mathcal{E}_\tm &=& \iint \frac{d^2 k_\perp}{(2\pi)^2} \int_0^\infty \frac{d\lambda}{\kappa} \alpha(\lambda) \left(\lambda^2 - 2 \kappa^2 \right)r^\tm(\lambda,\kappa) e^{-2a \kappa },\ann
\mathcal{E}_\te &=& \iint \frac{d^2 k_\perp}{(2\pi)^2} \int_0^\infty \frac{d \lambda}{\kappa} \alpha(\lambda) \lambda^2 r^\te (\lambda,\kappa) e^{-2a \kappa },\label{eq:ECP}
\end{eqnarray}
where $\kappa = \sqrt{k_\perp^2 + \lambda^2 }$ and $\alpha$ is polarizability of atom. 

To take into account temperature we have to change $\int_0^\infty d\lambda \to 2\pi T \sum_{n=0}^\infty{}'$ and $\lambda \to \xi_n\ (\kappa\to \kappa_n = \sqrt{k_\perp^2 + \xi_n^2})$, where $\xi_n = 2\pi n T$ being the Matsubara frequencies. We obtain the following expressions for free energy \cite{Khusnutdinov:2018:tccpinp2ddm}
\begin{eqnarray}
\mathcal{F}_\tm &=& \frac{T }{2\pi}\sum_{n=0}^\infty{\!}'\!\iint\! \frac{d^2 k_\perp}{\kappa_n}  \alpha_n \left(\xi_n^2 - 2 \kappa_n^2 \right) r^\tm(\xi_n,\kappa_n) e^{-2a \kappa_n },\ann
\mathcal{F}_\te &=& \frac{ T }{2\pi}\sum_{n=0}^\infty{\!}' \iint \frac{d^2 k_\perp}{\kappa_n} \alpha_n\xi_n^2  r^\te(\xi_n,\kappa_n) e^{-2a \kappa_n },\label{eq:Fa}
\end{eqnarray}
where $\alpha_n = \alpha (\xi_n)$. The ideal case appears by formal limit $\eta^{\te,\tm}_n \to\infty\ (r^\te \to -1,r^\tm \to 1)$. 

Taking into account the Poisson summation formula (see, for example, Ref. \cite{Bordag:2009:ACE}) we obtain following expression for free energy
\begin{eqnarray}
	\frac{\mathcal{F}_\tm}{\mathcal{E}_\CP} &=& -\frac{8}{3}\sum_{l=0}^\infty{}'  \int_{0}^\infty z^3 dz \int_0^1 dx 
	\cos \left(\frac{z x l}{aT}\right)   \ann
	&\times& \frac{\alpha (\lambda)}{\alpha(0)} \left(x^2 -2\right) r^\tm(x,1) e^{-2z },\ann
	\frac{\mathcal{F}_\te}{\mathcal{E}_\CP} &=& -\frac{8}{3}\sum_{l=0}^\infty{}'  \int_{0}^\infty z^3  dz \int_0^1 dx 
	\cos \left(\frac{z x l}{aT}\right)    \ann
	&\times& \frac{\alpha (\lambda)}{\alpha(0)}x^2 r^\te(x,1) e^{-2z },\label{eq:FPoi}
\end{eqnarray}
normalized to the $\mathcal{E}_{\CP} = - 3\alpha (0)/8\pi a^4$ -- Casimir-Polder energy for ideal plane/atom. Here $\lambda = \frac{zx}{a},\ k = \frac{z}{a}\sqrt{1-x^2}$ and the prime means factor $1/2$ for $l=0$. This form is more suitable for analysis at low temperature. Zero terms, $l=0$, in \eqref{eq:FPoi} coincides exactly with that obtained for zero temperature in Ref.  \cite{Khusnutdinov:2016:CPefasocp,*Kashapov:2016:TCefpls} (see Eq. \eqref{eq:ECP}) but with temperature and chemical potential dependence through  the conductivity. We extract the zero term 
\begin{equation}
\mathcal{F} = \mathcal{F}_\tm + \mathcal{F}_\te = \mathcal{F}_0 + \Delta \mathcal{F}  \label{eq:CPT}
\end{equation}
and consider low temperature expansion for $\Delta \mathcal{F}$ and $ \mathcal{F}_0$ separately. 

\section{The low temperature expansions} \label{Sec:LowT}

To analyze $\Delta \mathcal{F}$ \eqref{eq:CPT} we use Erd\'elyi's lemmas for asymptotic expansion  integrals  \cite{Fedoryuk:1977:TSPM}. For completeness we reproduce them below. 

\textbf{Lemma 1} 
\begin{equation}
\int_0^a x^{\beta-1} f(x) e^{i\Lambda x} dx = \sum_{n=0}^{\infty} a_n \Lambda^{-(n+\beta)},\label{eq:Erd1}
\end{equation}
where 
\begin{equation*}
a_n = f^{(n)}(0)\frac{\Gamma \left( n+\beta\right) }{n!}  e^{\frac{i\pi}{2} (n+\beta)}. 
\end{equation*}

\textbf{Lemma 2} 
\begin{equation}
\int_0^a x^{\beta-1} f(x) \ln x e^{i\Lambda x} dx = \sum_{n=0}^{\infty} b_n(\Lambda) \Lambda^{-(n+\beta)},\label{eq:Erd2}
\end{equation}
where 
\begin{gather*}
b_n = f^{(n)}(0) \frac{\Gamma \left(n+\beta\right)}{n!}  e^{\frac{i\pi}{2} (n+\beta)}\ann
\times \left[-\ln\Lambda + \psi \left(n+\beta\right) + \frac{i\pi}{2}\right] , 
\end{gather*}
and $\psi (x) = \Gamma' (x)/\Gamma (x)$. Both Lemmas are valid as $\Lambda \to \infty$,  $f^{(n)}(a) =0$ and $\beta >0$. 

The free Casimir-Polder energy $\Delta \mathcal{F}$ maybe represented in the following form 
\begin{equation}
 \frac{\Delta \mathcal{F}}{\mathcal{E}_{\CP}} = \frac{8}{3}\Re \sum_{l=1}^\infty \int_{0}^\infty dz e^{i\Lambda z} \left(Y_\tm  + Y_\te \right), \label{eq:ECaCP}
\end{equation}
where $\Lambda = \frac{l}{a T}$ and 
\begin{eqnarray}
Y_\tm(z) &=& \frac{\alpha (\lambda)}{\alpha (0)} \int_{z}^{\infty} \frac{e^{-2s} s \left(2s^2 - z^2\right)}{s + z/\eta_\tm}ds, \ann
Y_\te(z) &=& \frac{\alpha (\lambda)}{\alpha (0)} \int_{z}^{\infty} \frac{e^{-2s} z^3 }{z + s/\eta_\te}ds.\label{eq:YnCP} 
\end{eqnarray}
Here $\lambda = z/a$ and $k = \sqrt{s^2 - z^2}/a$ are used for frequency and wave-vector, correspondingly. 

First of all, let us consider the case without spatial dispersion, $\eta = \eta (\lambda)$. Straightforward integration in Eq. \eqref{eq:YnCP} gives
\begin{eqnarray}
Y_\tm &=&  \frac{\alpha e^{-2z}}{2\alpha(0) \eta^3_\tm}  \left\{ \eta_\tm\left(2z^2 - z(1+2z) \eta_\tm + (1+z)^2 \eta^2_\tm\right)\right.\ann
&+& \left. 2z^3 (\eta^2_\tm -2) e^{2z \left(1+ \eta^{-1}_\tm \right)} \Gamma \left(0,2z \left(1+ \eta^{-1}_\tm\right)\right)\right\},\ann
Y_\te &=& \frac{\alpha \eta_\te}{\alpha(0)} z^3 e^{2z\eta_\te} \Gamma \left(0,2z (1+\eta_\te)\right) ,\label{eq:YExact}
\end{eqnarray}
where $\Gamma (a,b)$ is incomplete gamma function.  

Expansion at point $z=0$ contents logarithmic contribution 
\begin{equation}
 Y_{\tm,\te}  = \sum_{m= 0}^\infty A_m^{\tm,\te}z^m + \ln z \sum_{m= 3}^\infty B_m^{\tm,\te} z^m.\label{eq:Y}
\end{equation}
Taking into account this expansion and Lemmas we obtain expansion up to 4th power of $T$ for the energy at low temperature
\begin{gather}
\frac{\Delta\mathcal{F}_{\tm,\te}}{\mathcal{E}_{\CP}} = - \frac{\chi^2}{9}A_1^{\tm,\te} +  \frac{\chi^4}{90}\left\{A_3^{\tm,\te} \right.\ann
+\left. B_3^{\tm,\te} \left(\ln\frac{\chi}{2\pi} - \gamma_E + \frac{90\zeta'_R(4)}{\pi^4} + \frac{11}{6}  \right)\right\},\label{eq:ET0}   
\end{gather}
where $\chi = 2\pi a T$ and $\zeta_R(s)$ is Riemann zeta-function. 

From Eqs. \eqref{eq:YExact} we obtain in manifest form 
\begin{eqnarray}
A_1^\tm &=& - \frac{1}{2\eta_\tm}, \ A_1^\te = 0, \ann 
A_3^\tm &=& -\frac{1}{4 a^2 \eta_\tm} \left(\frac{\alpha ''}{\alpha} - \frac{\eta ''_\tm}{\eta_\tm} + \frac{2 \eta '^2_\tm}{\eta ^2_\tm} \right)  - \frac{2\eta'_\tm}{a \eta^3_\tm}\ann
&-&\frac{2}{\eta ^2_\tm} + \frac{1}{\eta_\tm } + \frac{1}{3} + \left(\gamma_E + \ln\left[\frac{2(1+\eta_\tm)}{\eta_\tm} \right] \right)B_3^\tm,\ann
A_3^\te &=& \left(\gamma_E + \ln \left[2(1+\eta_\te)\right]\right)B_3⁝^\te,\ann
B_3^\tm &=& \frac{2}{\eta^3_\tm} - \frac{1}{\eta_\tm},\ B_3^\te = -\eta_\te.
\end{eqnarray}
Taking into account these expressions we obtain asymptotic expansion of free \CP\ energy
\begin{eqnarray}
\frac{\Delta\mathcal{F}_\tm}{\mathcal{E}_{\CP}} &=& \frac{\chi^2}{18\eta_\tm} + \frac{\chi^4}{270}  \left\{1 - \frac{6}{\eta^2_\tm}\left(1 + \frac{\eta'_\tm}{a\eta_\tm}\right) \right.\ann
&-& \left.\frac{1}{2\eta_\tm} \left(5 + \frac{540 \zeta'_R(4)}{\pi^4}  + 6 \ln \left[\chi \frac{1+\eta_\tm }{\pi \eta_\tm}\right]\right.\right.\ann
&+& \left.\left.\frac{3}{2a^2} \left( \frac{\alpha''}{\alpha}  -  \frac{\eta''_\tm}{\eta_\tm} + \frac{2\eta'^2_\tm}{\eta^2_\tm} \right) \right)\right.\ann
&+& \left. \frac{6}{\eta^3_\tm} \left(\frac{11}{6} +  \ln \left[\chi \frac{1+\eta_\tm }{\pi \eta_\tm}\right] + \frac{90 \zeta'_R(4)}{\pi^4}\right) \right\},\ann
\frac{\Delta \mathcal{F}_\te}{\mathcal{E}_{\CP}} &=& - \frac{\chi^4 \eta_\te}{90} \left\{\frac{11}{6} + \ln \left[\chi \frac{1+\eta_\te}\pi\right] + \frac{90 \zeta'_R (4)}{\pi^4} \right\},\label{eq:Iap}
\end{eqnarray}
where functions $\alpha$ and $\eta$ and their derivatives are considered at zero argument. 

In ideal case \cite{Bezerra:2008:Ltaiatqr} the free energy for low temperatures has first correction $\sim \chi^4$
\begin{equation}
\left.\frac{\mathcal{F}}{\mathcal{E}_{\CP}}\right|_{T\to 0} = 1 - \frac{\chi^4}{135}.\label{eq:ideal}
\end{equation}

One comment is in order. Above expansions are valid if arguments of incomplete gamma functions in Eqs. \eqref{eq:YExact} are small, that is for $\chi \ll \frac{\eta_\tm}{1+\eta_\tm}$ and $\chi \ll \frac{1}{1+\eta_\te}$. Therefore, we may take limit to ideal case only for \TM\ mode in Eq. \eqref{eq:Iap}. To consider ideal case for \TE\ mode we have to take limit $\eta_\te \to \infty$ first of all in Eq. \eqref{eq:YExact} and then make expansion over $z$.  

The main term of expansion $\chi^2/18\eta_\tm$ is the same for $k\not = 0$. Indeed, to obtain $A_1^\tm$ we may take derivative of Eq. \eqref{eq:YnCP} with respect of $z$ and then take limit $z\to 0$. By proceed that way we obtain the same form of main term where $\eta_\tm$ is calculated for $\lambda=k=0$.   

Therefore, we observe that for all models of conductivities the main term of low temperature expansion proportional to $\chi^2$. 

Let us consider now zero term in Poisson representation $\mathcal{F}_0 = \mathcal{F}^0_\tm + \mathcal{F}^0_\tm$ 
\begin{eqnarray}
	\frac{\mathcal{F}^0_\tm}{\mathcal{E}_\CP} &=& -\frac{4}{3}\int_{0}^\infty z^3 dz \int_0^1 dx \frac{\alpha (\lambda)}{\alpha(0)} \left(x^2 -2\right) r^\tm(x,1) e^{-2z },\ann
	\frac{\mathcal{F}^0_\te}{\mathcal{E}_\CP} &=& -\frac{4}{3}\int_{0}^\infty z^3  dz \int_0^1 dx \frac{\alpha (\lambda)}{\alpha(0)}x^2 r^\te(x,1) e^{-2z }. \label{eq:ZeroP}
\end{eqnarray}
As noted above it coincides exactly with that obtained for zero temperature in Ref. \cite{Khusnutdinov:2016:CPefasocp,*Kashapov:2016:TCefpls}, but with additional dependence on the temperature and chemical potential through  dependence of conductivity on these parameters (see Sec. \ref{Sec:PT}). These expressions tend to $1/2$ for ideal ($\eta \to \infty$) case and for $a\to\infty$. 

\section{Numerical analysis} \label{Sec:Numeric}

Let us compare numerically the formulas obtained \eqref{eq:Iap} with exact numerical calculations for different models of conductivity. Let us denote for simplicity $\delta \mathcal{F}_n = \Delta \mathcal{F}/\mathcal{E}_{\CP}$ where $n=0$ corresponds to exact expression  calculated numerically and $n=1,2$ corresponds to first ($\sim \chi^2$) and second ($\sim \chi^4$) approximations in \eqref{eq:Iap}. Numerically we use \eqref{eq:Fa} and subtract \eqref{eq:ZeroP}. To estimate error we plot relative error function $E_n = (\delta \mathcal{F}_n - \delta \mathcal{F}_0)/\delta \mathcal{F}_0 \cdot 100\%$. 

For definiteness we consider Hydrogen atom in framework of one-oscillator model (see Ref. \cite{Khusnutdinov:2016:CPefasocp}) and distance $a=10nm$ between atom and plane of graphene. Then the interval of temperatures $T \in [0, 100^\circ] K $ corresponds to interval of parameter $\chi \in [0,2.7]\cdot 10^{-3}$. Different models of graphene's conductivity briefly discussed in Appendix \ref{Sec:Models}.  

\subsection{Constant conductivity model}

\begin{figure}[ht]
\includegraphics[width=4.5 truecm]{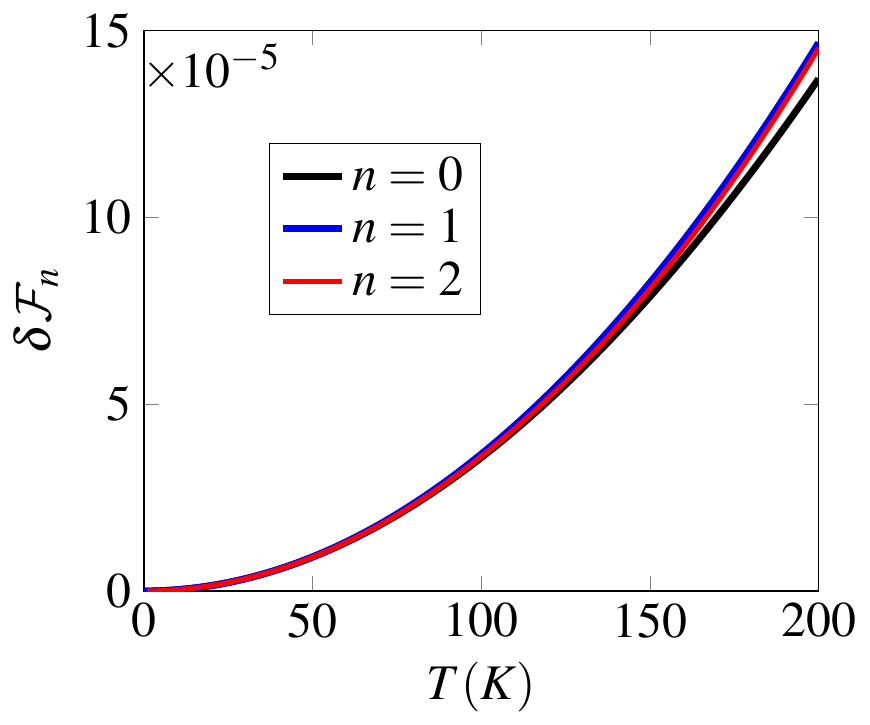}\includegraphics[width=4.35 truecm]{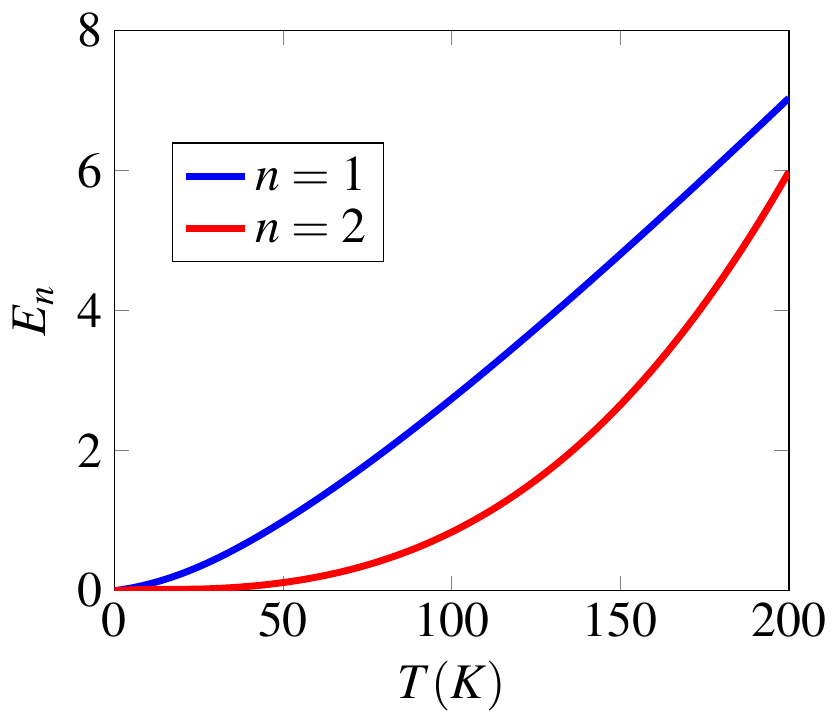}
\caption{Constant conductivity model of graphene $\eta_\tm = \eta_\te = 0.0114$ and Hydrogen atom at distance $a=10nm$ from graphene. Left panel: $\delta \mathcal{F}_n = \Delta \mathcal{F}/\mathcal{E}_\CP$. $n=0$ is numeric simulation, $n=1$ is approximation up to $T^2$, $n=2$ is approximation up to $T^4$. Right panel: relative error in percents. } \label{fig:1}
\end{figure}

First model is constant conductivity model. For graphene we use universal conductivity, $\sigma_{gr} = e^2/4\hbar$ and then $\eta_\tm = \eta_\te = \eta_{gr} = 0.0114$. Fig. \ref{fig:1} illustrates numerical evaluation exact expression and approximations obtained and relative error. 
We observe that relative error for second approximation is not more then $1\%$ up to $T=100K$. 

\subsection{Drude-Lorentz model}

In the case of Drude-Lorentz 7-oscillator model of conductivity agreement is not so good. The point is that even the conductivity at zero frequency equals to graphene universal conductivity, but derivatives of first and second order are very huge. The contributions to $\chi^4$ contain terms 
\begin{equation}
 \eta = 0.0114,\  \frac{\eta'}{a\eta} = 8.76\cdot 10^3,\ \frac{\eta''}{a^2\eta} = -1.85\cdot 10^7,
\end{equation}   
and the ratio of second term $\sim \chi^4$ and first $\sim \chi^2$ is $0.67$ for $T=10^\circ K$ and $67$ for $T=100^\circ K$. 
\begin{figure}[ht]
\includegraphics[width=4.35 truecm]{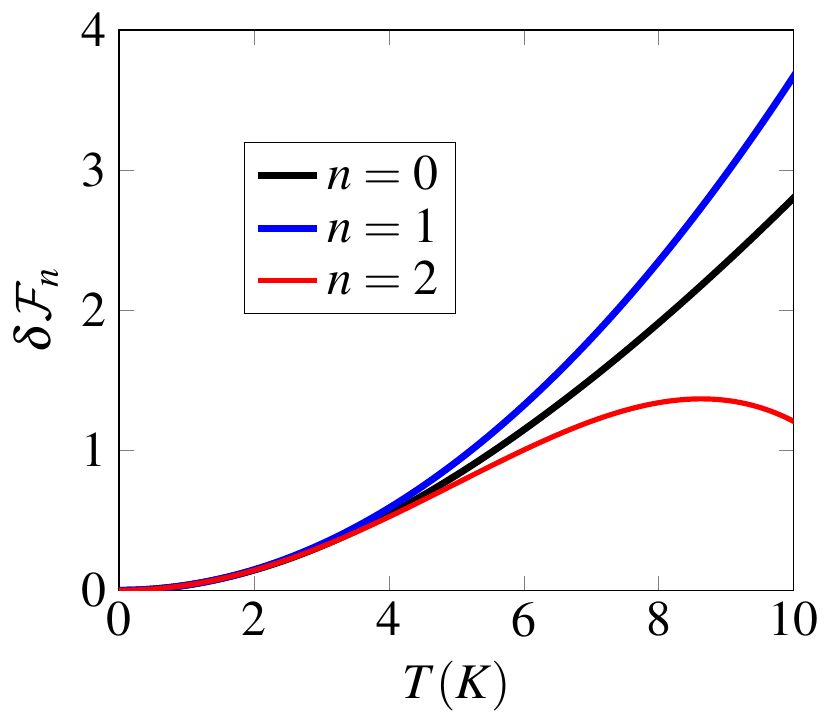}\includegraphics[width=4.5 truecm]{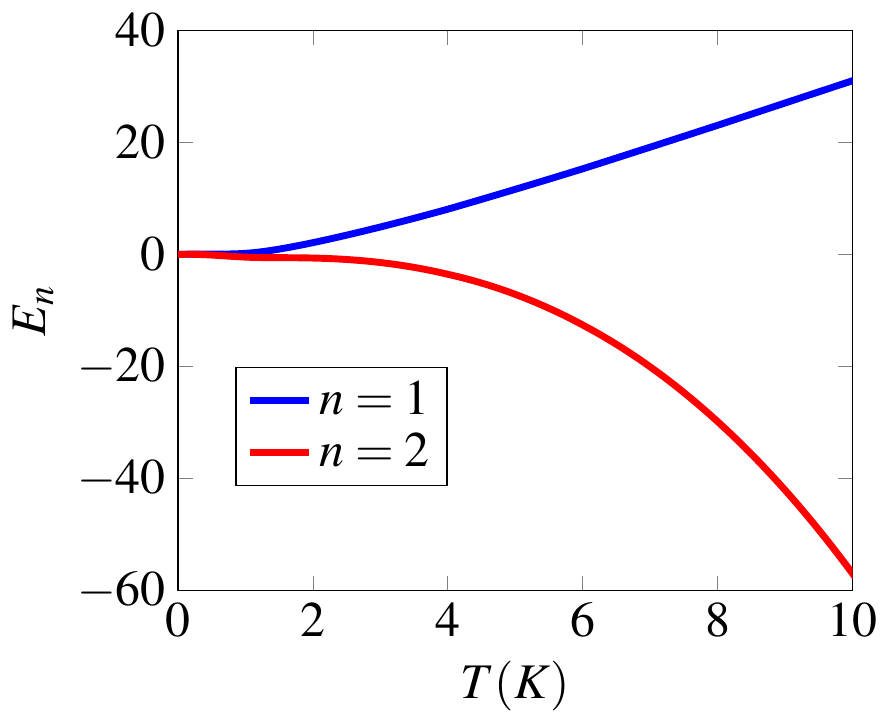}
\caption{The Drude-Lorentz model of graphene's conductivity and Hydrogen atom at distance $a=10nm$ from graphene. Left panel: $\delta \mathcal{F}_n = \Delta \mathcal{F}/\mathcal{E}_\CP$. $n=0$ is numeric simulation, $n=1$ is approximation up to $T^2$, $n=2$ is approximation up to $T^4$. Right panel: relative error in percents. } \label{fig:2}
\end{figure}
For this reason the low temperature expansion is valid for very low temperature (see Fig. \ref{fig:2}).   

\subsection{Polarization tensor approach} 

The last model is polarization tensor model of conductivity developed in Refs.  \cite{Bordag:2009:CibapcagdbtDm,*Fialkovsky:2011:FCefg,*Bordag:2016:ECefdg,*Bordag:2017:EECefdg}. In this case the conductivity depends on frequency, wave-vector, temperature and chemical potential $\eta_{\tm,\te}(\lambda,k,T,\mu)$. In the case under consideration, $k=0$,  
\begin{gather}
\frac{\eta_{\tm,\te}}{\eta_{gr}} = \frac{4m}{\pi \lambda}\left\{1 + \frac{\left(\frac{\lambda}{2m}\right)^2 -1}{\left(\frac{\lambda}{2m}\right)}  \arctan \left(\frac{\lambda}{2m}\right)\right\}\ann
+ \frac{16 }{\pi \lambda} \int_m^\infty dz \frac{z^2 + m^2}{4z^2 +\lambda^2}\left\{\frac{1}{e^{\frac{z+\mu}{T}}+1}  + \frac{1}{e^{\frac{z-\mu}{T}}+1}\right\} . \label{eq:sigmak0}
\end{gather}
\begin{figure}[ht]
\includegraphics[width=4.5 truecm]{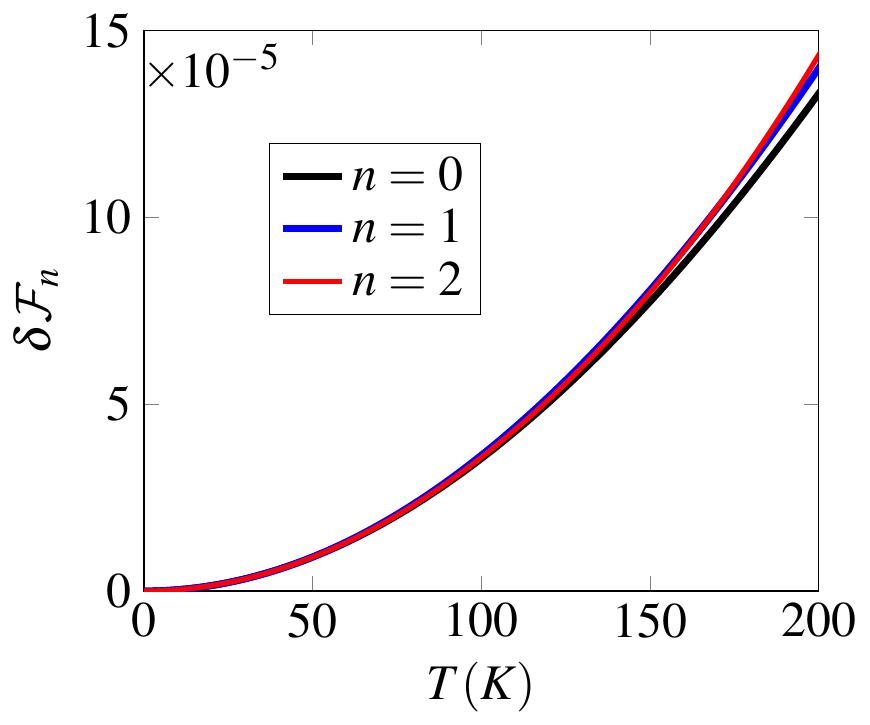}\includegraphics[width=4.35 truecm]{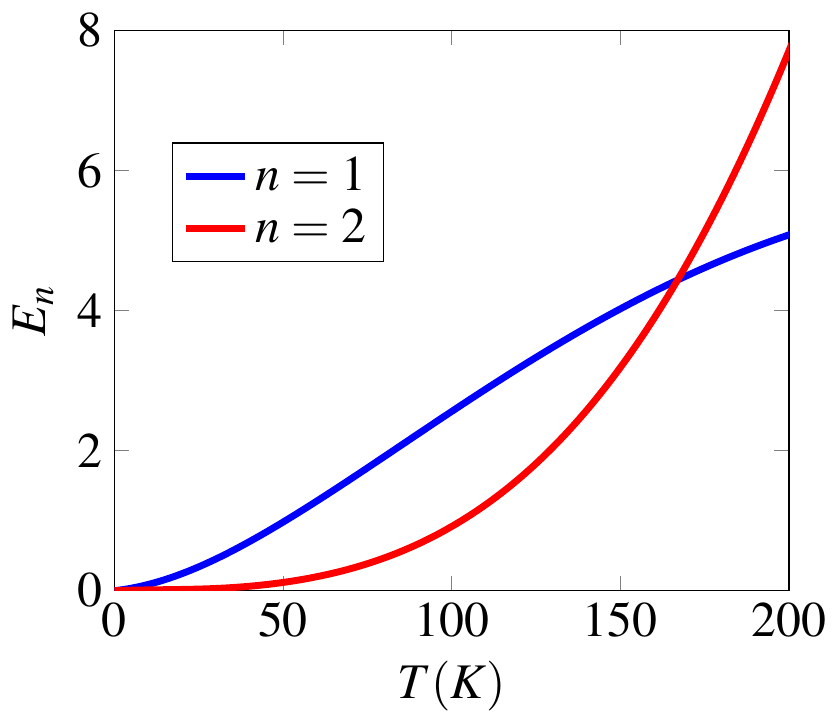}
\caption{Polarization tensor approach for conductivity with $m=\mu= 0$ and $\gamma = 0.1eV$ at distance $a=10nm$ from graphene. Left panel: $\delta \mathcal{F}_n = \Delta \mathcal{F}/\mathcal{E}_\CP$. $n=0$ is numeric simulation, $n=1$ is approximation up to $T^2$, $n=2$ is approximation up to $T^4$. Right panel: relative error in percents. } \label{fig:3}
\end{figure}
In the static limit, $\lambda\to 0$, the conductivity is divergent (but transmission and reflection coefficients \eqref{eq:transmission} tend to that for ideal case). To make it finite we cut $\lambda$ on minimal value $\gamma$. The threshold parameter $\gamma$ appears in natural way in framework of Kubo approach calculation of conductivity \cite{Falkovsky:2007:Sdogc} as a scattering rate. 

\begin{figure}[ht]
\includegraphics[width=4.5 truecm]{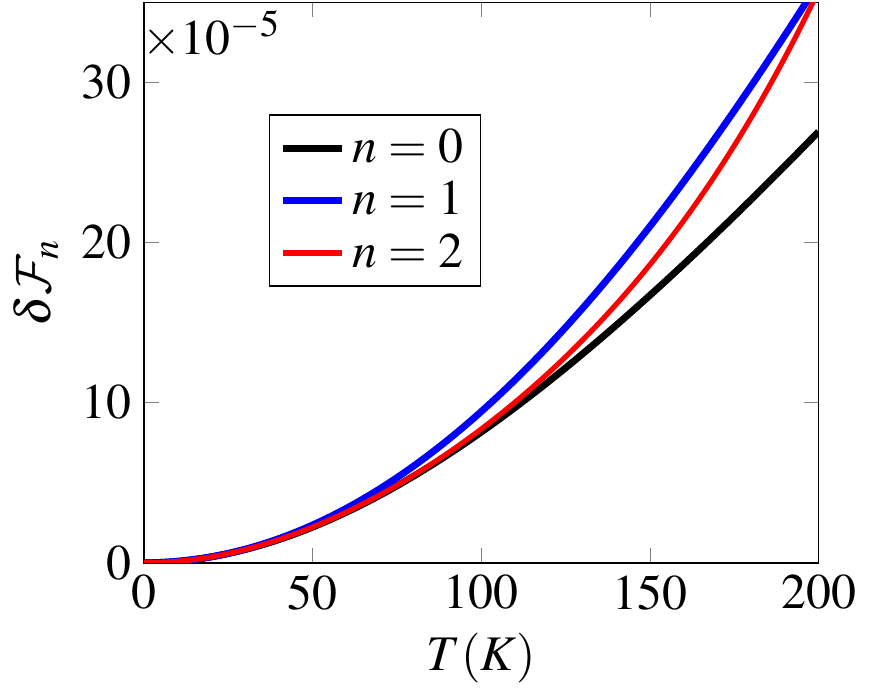}\includegraphics[width=4.35 truecm]{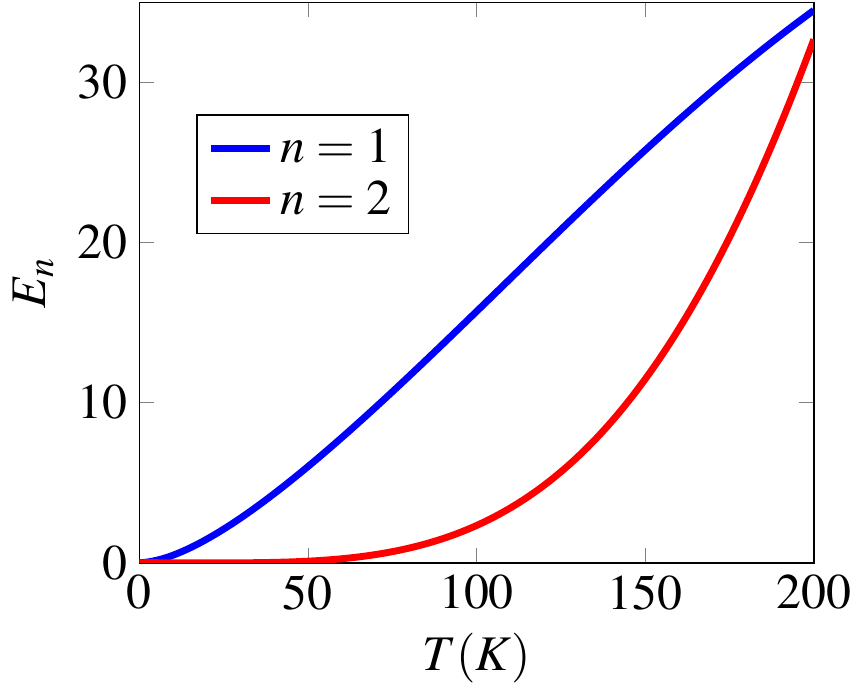}
\caption{Polarization tensor approach for conductivity with $m=0.1eV, \mu= 0.05eV$ and $\gamma = 0.1eV$ at distance $a=10nm$ from graphene. Left panel: $\delta \mathcal{F}_n = \Delta \mathcal{F}/\mathcal{E}_\CP$. $n=0$ is numeric simulation, $n=1$ is approximation up to $T^2$, $n=2$ is approximation up to $T^4$. Right panel: relative error in percents. } \label{fig:4}
\end{figure}

Numerical evaluations (see Figs. \ref{fig:3}, \ref{fig:4}) demonstrate good agreement expansion \eqref{eq:Iap} with exact expression \eqref{eq:Fa}. 

Let us consider now zero temperature term $\mathcal{F}_0$ given by Eqs. \eqref{eq:CPT}, \eqref{eq:ZeroP}. In the framework of the model under consideration it depends on the temperature and chemical potential. Expansion of conductivity over temperature is given in Appendix \ref{Sec:AppB}. 

For $m=\mu=0$ the first correction $\sim T^3$ (see Eq. \eqref{eq:apBzeros}) and reads 
\begin{equation}
\eta_\te = \eta_\tm = \eta_{gr} + \frac{48}{\pi} \zeta_R(3) \frac{T^3}{\lambda^3}.
\end{equation}
Therefore, 
\begin{equation}
 \frac{\mathcal{F}_0}{\mathcal{E}_\CP} = \frac{\mathcal{F}_0^{T=0}}{\mathcal{E}_\CP} + T^3 \beta \label{eq:beta}
\end{equation}
where
\begin{eqnarray}
	\beta &=& \frac{64 \zeta_R(3)\eta_{gr}}{\pi}\int_{0}^\infty z^3 dz \int_0^1 dx \frac{\alpha (\lambda)}{\alpha(0)} \frac{x e^{-2z}}{\lambda^3} \ann
	&\times& \left\{\frac{2- x^2}{(x + \eta_{gr})^2} +  \frac{x^2}{(1+ x\eta_{gr})^2}\right\}.
\end{eqnarray}
For $a=10nm$ and $\gamma=0.1 eV$ we have $\beta = 2.33\cdot 10^{-12}$ where $T(K)$. 

For $m=0.1eV$ and $\mu = 0.05eV$ the first correction is exponentially small $\sim e^{-\frac{m-\mu}{T}}$ (see Eq. \eqref{eq:apBnonzero}). 
\begin{figure}[ht]
\includegraphics[width=4.5 truecm]{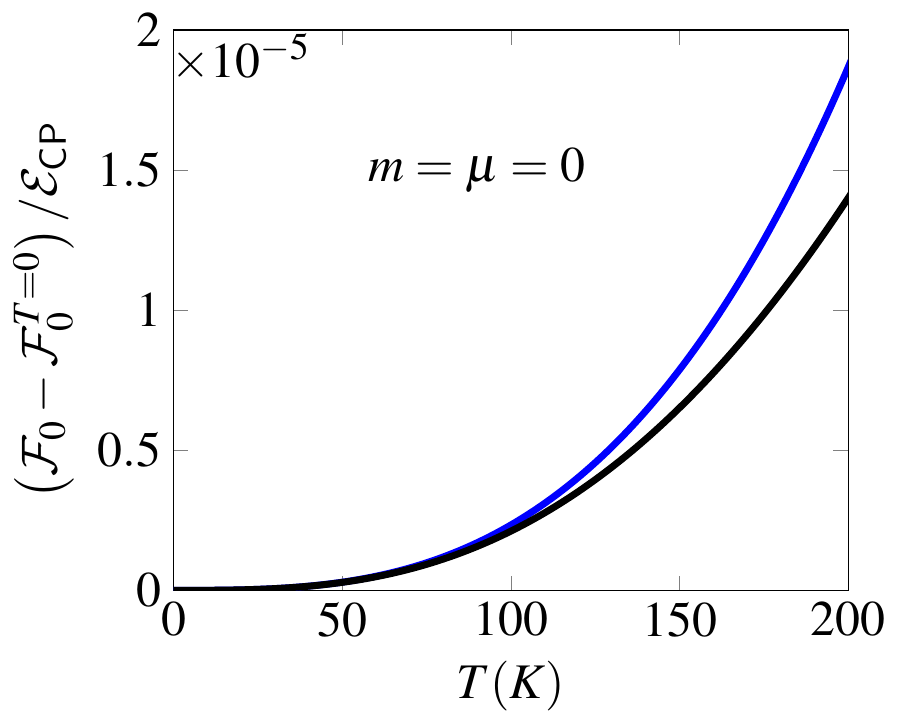}\includegraphics[width=4.1 truecm]{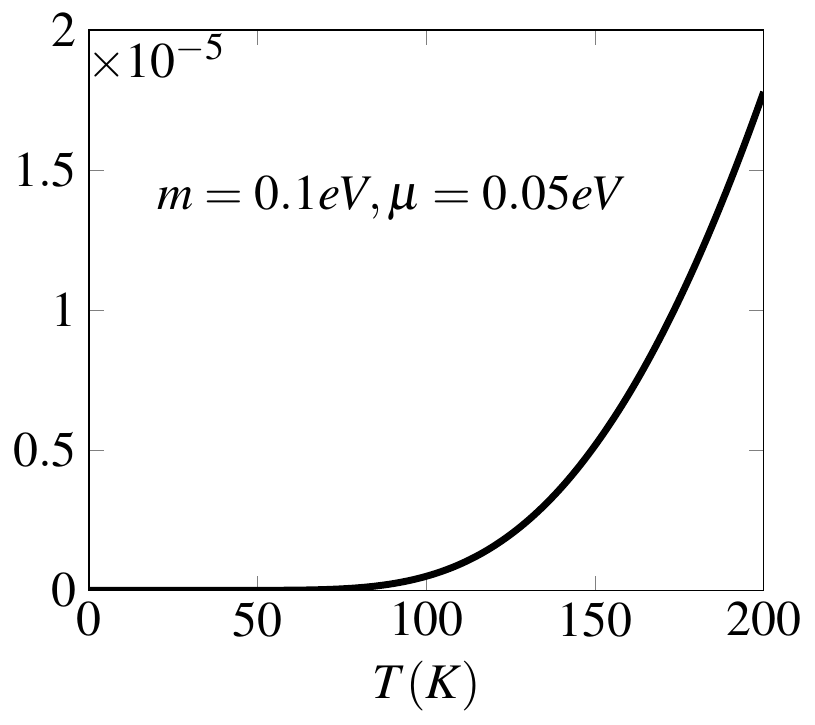}
\caption{Polarization tensor approach for conductivity. Zero term $\left(\mathcal{F}_0 - \mathcal{F}_0^{T=0} \right) / \mathcal{E}_{\textsf{CP}}$. Left panel: $m=\mu=0$ and $\gamma = 0.1eV$. Black curve is exact expression \eqref{eq:ZeroP} and blue is first approximation \eqref{eq:beta}. Right panel: $m=0.1eV,\mu=0.05eV$ and $\gamma=0.1eV$. Temperature contribution is exponentially small up $100^\circ K$.} \label{fig:5}
\end{figure}
Fig. \ref{fig:5} illustrates temperature correction of zero temperature term due to conductivity dependence of temperature and chemical potential.  

\section{Conclusion} \label{Sec:Conclusion}

We have obtained the analytic expression of low temperature expansion of the Casimir-Polder (van der Waals) energy for a system which contains an atom and conductive plane. The conductivity is characterized by symmetric $2D$ conductivity tensor. The eigenvalues of this tensor are conductivity of \TE\ and \TM\ modes. The main term of expansion $\sim T^2$ comes from \TM\ mode, the next terms $\sim T^4$ and $\sim T^4 \ln T$. Numerical analysis shows good agreement expansion obtained with exact expressions.  

\begin{acknowledgements}
NK was supported in part by the Russian Foundation for Basic Research Grant No. 16-02-00415-a and by the grants 2016/03319-6 and 2017/50294-1 of S\~ao Paulo Research Foundation (FAPESP).
\end{acknowledgements}

\appendix

\section{Models of conductivity}\label{Sec:Models}
In this appendix we consider different models of graphene conductivity which were used for numerical analysis in Sec. \ref{Sec:Numeric}. We normalize conductivity to the universal conductivity of graphene $\sigma_{gr} = e^2/4\hbar$ and mark these quantities by overline.  

\subsection{Constant conductivity}

It is well-known that the graphene conductivity is a constant ($\sigma_{gr} = e^2/4\hbar$) over a relatively large frequency range, near infrared to optical 
\cite{Gusynin:2007:Mcig,*Nair:2008:FSCDVToG}. 
For this reason we consider the model in which the conductivity equals to this value for whole frequencies. In this case
 \begin{equation}
 \overline{\bm{\sigma}} = \bm{I},
 \end{equation} 
and $\overline{\sigma}_\te = \overline{\sigma}_\tm = 1$. This approximation of conductivity was used intensively in Ref. \cite{Khusnutdinov:2016:CPefasocp}.

\subsection{Drude-Lorents model}

We use conductivity of graphite alongside to planes which is approximated with high precision by Drude-Lorentz model consisting of a Drude term and seven Lorentz oscillators according to \cite{Djurisic:1999:Opog}:
\begin{equation}
\sigma (\omega) = \frac{f_0 \omega_p^2}{\widetilde{\gamma}_0 - i\omega} + \sum_{j =1}^7\frac{i \omega f_j \omega_p^2}{ \omega^2 - \omega_j^2 + i\omega \widetilde{\gamma}_j }.\label{eq:DL}
\end{equation}
We multiply it on the length scale $d=0.2245nm$ which is close to interplane distance of graphite $d_{gr} = 0.3345nm$ and obtain  $\overline{\sigma}_\te = \overline{\sigma}_\tm = \overline{ \sigma}$ and  
 \begin{equation}
\overline{\bm{\sigma}} =\overline{ \sigma} \bm{I},
\end{equation} 
where 
\begin{equation}
\overline{\sigma}(\lambda)   = \frac{\overline{\sigma}_0 \gamma_0}{\gamma_0 + \lambda} + \sum_{j = 1}^7\frac{\lambda \overline{\sigma}_j \gamma_j }{ \lambda^2 + \lambda_j^2 + \lambda \gamma_j }.\label{eq:DLcompl}
\end{equation}
With scale $d=0.2245nm$ we obtain right limit for small frequencies $\overline{ \sigma}(0) = 1$.  Here, $\gamma_j$ is the relaxation time and $\omega_j$ is the characteristic frequency for the $j$-th term. All parameters of of this model maybe found in Ref. \cite{Khusnutdinov:2016:CPefasocp}.

\subsection{Polarization tensor approach}\label{Sec:PT}

In Ref. \cite{Bordag:2009:CibapcagdbtDm,*Fialkovsky:2011:FCefg,*Bordag:2016:ECefdg,*Bordag:2017:EECefdg} was used relation between $(2+1)D$ polarization tensor, $\Pi_{\mu\nu}$ and tensor of conductivity, $\sigma_{lj}$,  
\begin{equation}
\sigma_{ln} = \frac{\Pi_{ln}}{\ii\omega},
\end{equation}
and obtained polarization tensor in general form, which depends on $\lambda,k,T,\mu$. Due to gauge invariance the polarization tensor has only two independent components, for example, $\Pi_{00}$ and $\Pi_{tr} = \Pi^{\mu\nu} \gs_{\mu\nu} = \Pi_{00} - \Pi_{11} - \Pi_{22}$. 

In Ref. \cite{Bordag:2016:ECefdg,*Bordag:2017:EECefdg} has obtained elegant representation of polarization tensor. Taking into account these expressions and expressions \eqref{eq:eigen} for \TE\ and \TM\ conductivities we have  
\begin{equation}
\overline{\sigma}_\te = \frac{4}{e^2\lambda}\left(\Pi_{tr} -\frac{\lambda^2 + k^2}{k^2} \Pi_{00}\right),\ \overline{\sigma}_\tm = \frac{4\lambda}{e^2k^2} \Pi_{00}.
\end{equation}

According with Ref. \cite{Bordag:2016:ECefdg,*Bordag:2017:EECefdg} we divide conductivity into two parts  
\begin{equation}
\overline{\sigma}_{\tm,\te} = \overline{\sigma}_{\tm,\te}^0 + \Delta\overline{\sigma}_{\tm,\te},
\end{equation}
where the first term does not depend on the temperature, $T$ and chemical potential, $\mu$ and read
\begin{equation}
\overline{\sigma}_\te^0 = \frac{4m}{\pi \lambda} \Psi \left(\frac{k_F}{2m}\right),\ \overline{\sigma}_\tm^0 = \frac{4m \lambda}{\pi k_F^2} \Psi \left(\frac{k_F}{2m}\right), \label{eq:zero}
\end{equation}
with $\Psi(x) = 1 + \frac{x^2 - 1}{x} \arctan x$ and $k_F = \sqrt{\lambda^2 + v_F^2 k^2}$. The correction $\Delta\overline{\sigma}_{\tm,\te}$ maybe represented in the following form 
\begin{widetext}
\begin{equation*}
\Delta\overline{\sigma}_\te = \frac{8}{\pi \lambda} \Re \int_m^\infty dz \frac{(4m^2+q^2) (q^2k_F^2 + 4 m^2 k^2 v_F^2) - q^2 k_F^2\lambda^2 }{r(q^2k_F^2 + 4m^2 k^2 v_F^2 + q \lambda r)} \Theta , \Delta\overline{\sigma}_\tm = \frac{8}{\pi} \Re \int_m^\infty dz \frac{q (q^2 +k^2 v_F^2 + 4m^2) - \lambda r  }{r(r + q \lambda)} \Theta,\label{eq:Sigma}
\end{equation*}
\end{widetext}
where $\Theta = (e^{\frac{z + \mu}{T}} +1)^{-1} + (e^{\frac{z - \mu}{T}} +1)^{-1}$, $r = \sqrt{k_F^2 (q^2 + k^2 v_F^2) + 4m^2 k^2 v_F^2}$, and $q=\lambda-2i z$. 

In the gapless case, $m=0$, we obtain a little bit simple expressions
 \begin{eqnarray}
 	\Delta\overline{\sigma}_\te &=& \frac{8}{\pi \lambda} \Re \int_0^\infty dz \frac{q\left(q^2  - \lambda^2 \right)\Theta}{r_0 (qk_F + \lambda r_0)} ,\ann 
 	\Delta\overline{\sigma}_\tm &=& \frac{8}{\pi k_F} \Re \int_0^\infty dz \frac{\left(q r_0 - \lambda k_F\right) \Theta}{k_F r_0 + q \lambda},\label{eq:Sigma1}
\end{eqnarray}
with $r_0 = \sqrt{q^2 + k^2 v_F^2}$ and $\overline{\sigma}_\te^0 = k_F/\lambda,\ \overline{\sigma}_\tm^0 = \lambda /k_F$. Let us consider some special limits. 

\subsubsection{$k\to 0$}

In this limit we obtain $\overline{\sigma}_\te = \overline{\sigma}_\tm$ and 
\begin{equation}
\Delta\overline{\sigma}_\te = \Delta\overline{\sigma}_\tm = \frac{16 }{\pi \lambda} \int_m^\infty dz \frac{z^2 + m^2}{4z^2 +\lambda^2} \Theta. \label{eq:sigmak0}
\end{equation}
The zero temperature contribution reads 
\begin{equation}
\overline{\sigma}_\te^0 = \overline{\sigma}_\tm^0 =  \frac{4m}{\pi \lambda} \Psi \left(\frac{\lambda}{2m}\right).\label{eq:sigmak0-1}
\end{equation}

These expressions have peculiarity in static limit $\lambda\to 0$. Zero temperature terms, $\overline{\sigma}_{\te,\tm}^0 \to 0$  (for $m\not = 0$) and $\overline{\sigma}_{\te,\tm}^0 =1$  (for $m = 0$), but $\Delta\overline{\sigma}_{\te,\tm} \sim \lambda^{-1}$. 

For $m=0$ and $\mu=0$ this expression coincides with that obtained by method Kubo \cite{Falkovsky:2007:Sdogc} (see Eq. \eqref{eq:KuboS}) where scattering rate, $\gamma$, was used. For this reason we change $\lambda \to \lambda + \gamma$, that is cut $\lambda$ at minimal value $\gamma$ as was used in Ref. \cite{Falkovsky:2007:Sdogc}. 

\subsubsection{$T\to 0$} 

In this case we observe from Eq. \eqref{eq:Sigma} that temperature contribution in conductivity is zero if $\mu \leq m$ and it reads
\begin{widetext}
	\begin{eqnarray}
	\Delta\overline{\sigma}_\te &=& \frac{8}{\pi \lambda} \Re \int_m^\mu dz \frac{(4m^2+q^2) (q^2k_F^2 + 4 m^2 k^2 v_F^2) - q^2 k_F^2\lambda^2 }{r(q^2k_F^2 + 4m^2 k^2 v_F^2 + q \lambda r)} \theta (\mu-m) ,\ann
	\Delta\overline{\sigma}_\tm &=& \frac{8}{\pi} \Re \int_m^\mu dz \frac{q (q^2 +k^2 v_F^2 + 4m^2) - \lambda r }{r(r + q \lambda)} \theta (\mu-m),
	\end{eqnarray}
\end{widetext}
where $\theta (x)$ is step function. The zero terms have the same form \eqref{eq:zero}.  Therefore, we have additional contribution due to chemical potential. If $\mu \leq m$ the conductivity is defined by zero temperature contribution \eqref{eq:zero}.

In the case of zero mass gap, $m=0$, the conductivity is zero if $\mu \leq 0$ and it reads
\begin{eqnarray}
\Delta\overline{\sigma}_\te &=& \frac{8}{\pi \lambda} \Re \int_0^\mu dz \frac{q\left(q^2  - \lambda^2 \right)}{r_0 (qk_F + \lambda r_0)}\theta (\mu) ,\ann 
\Delta\overline{\sigma}_\tm &=& \frac{8}{\pi k_F} \Re \int_0^\mu dz \frac{q r_0 - \lambda k_F}{k_F r_0 + q \lambda}\theta (\mu).\label{eq:Sigma2}
\end{eqnarray}
The expansion over $T$ up to $T^4$ obtained in Appendix \ref{Sec:AppB}.

\subsection{Kubo approach}\label{Sec:Kubo}

The tensor of conductivity $\sigma_{ij}(\lambda,k,T)$ was obtained in Ref. \cite{Falkovsky:2007:Sdogc} in framework of Kubo approach. The eigenvalues in this case have the following form  
\begin{equation}
\overline{\sigma}_{\te \atop \tm} =  \int_0^\infty \hspace{-1ex}xdx \int_0^{2\pi} \hspace{-1.5ex}d\varphi \left\{ K_- {\sin^2\varphi  \atop \cos^2\varphi } + K_+ {\cos^2\varphi \atop \sin^2\varphi}  \right\} , 
\end{equation}
where
\begin{eqnarray*}
K_\mp &=& \frac{4}{\pi^2}\frac{\tanh \left(p\nu_+ \right) \mp \tanh \left(p\nu_-\right)}{(\nu_+ \mp \nu_-) (1 +  (\nu_+ \mp \nu_-)^2)},\ann
\nu_\pm &=& \sqrt{x^2 + \frac{b^2}{4} \pm b x \cos\varphi},
\end{eqnarray*}
and $\gamma$ is scattering rate. Three parameters $\lambda,k,T$ are combined in two dimensionless parameters 
\begin{equation}
p = \frac{\lambda + \gamma}{2T},\ b=\frac{k v_F}{\lambda + \gamma}.
\end{equation}

Let us consider different limits. 

\subsubsection{$k \to 0$}

We obtain that   
\begin{equation}
\overline{ \sigma}_\te = \overline{ \sigma}_\tm =  \frac{4\ln 2}{\pi p } + \frac{2}{\pi} \int_0^\infty \frac{\tanh \left(\frac{p x}{2} \right) }{x^2+1} dx. \label{eq:KuboS}
\end{equation}
The same expression maybe obtained from that in framework of polarization tensor approach \eqref{eq:sigmak0}, \eqref{eq:sigmak0-1} taking into account scattering rate $\gamma$.

The conductivity without spatial dispersion was also obtained in Ref. \cite{Gusynin:2007:Mcig}.  

\subsubsection{$T\to 0$}

The conductivities read
\begin{eqnarray}
\overline{ \sigma}_\te &=& - \frac{4}{\pi} \frac{2+b^2}{b^2} \arctan b + 2+ \frac{8}{\pi b} + \frac{4}{b^2} - \frac{4 + 3b^2}{b^2 \sqrt{1+b^2}},\ann
\overline{ \sigma}_\tm &=& -\overline{ \sigma}_\te - b + \frac{2+b^2}{\sqrt{1+b^2}}.
\end{eqnarray}
In the limit $b\to 0\ (k\to 0\ \textrm{or}\ \lambda \to \infty)$ we obtain $\overline{ \sigma}_\te = \overline{ \sigma}_\tm = 1$, that is $\overline{\bm{\sigma}} =  \bm{I}$.  

\section{Low temperature expansion of conductivity}\label{Sec:AppB}

The temperature correction for conductivity has the following form ($i=\te,\tm$)
\begin{equation}
\Delta\overline{\sigma}_i = \int_m^\infty dz f_i(z)\Theta,
\end{equation}
where 
\begin{eqnarray}
f_\te (z) &=& \frac{8}{\pi \lambda} \Re \frac{(4m^2+q^2) (q^2s_F^2 + 4 m^2 k^2 v_F^2) - q^2 s_F^2\lambda^2 }{r(q^2s_F^2 + 4m^2 k^2 v_F^2 + q \lambda r)},\ann
f_\tm (z) &=& \frac{8}{\pi} \Re \frac{q (q^2 +k^2 v_F^2 + 4m^2) - \lambda r  }{r(r + q \lambda)}. 
\end{eqnarray}

We have the following expansions in two domains ($m \geq 0$) for $T\to 0$:

I. $\mu \geq m$

\begin{eqnarray}
\Delta\overline{\sigma}_i &=& \int_m^{\mu} f_i (z) dz + \sum_{n=0}^\infty T^{n+1} f^{(n)}_i(\mu) (-1)^n \ann
&\times& \left\{ F_n\left[\frac{\mu + m}{T}\right] + F_n\left[\frac{\mu - m}{T}\right]\right.\ann
&-&\left. (1-(-1)^n) F_n[0]\right\}.
\end{eqnarray}

II. $\mu \leq m$

\begin{eqnarray}
\Delta\overline{\sigma}_i &=& \sum_{n=0}^\infty T^{n+1} f^{(n)}_i(\mu) \ann
&\times& \left\{ (-1)^nF_n\left[\frac{\mu + m}{T}\right] +   F_n\left[\frac{m - \mu}{T}\right]\right\}.
\end{eqnarray}
Here we used notation
\begin{equation}
 F_n[x] = \frac{1}{n!} \int_x^\infty \frac{z^n dz}{e^z +1}.
\end{equation}
This function has the following behavior at large and small argument
\begin{equation}
 F_n[x]_{x\to \infty} = e^{-x} \frac{x^n}{n!},\ F_n[0] = (1-2^{-n}) \zeta_R (n+1).
\end{equation}

The above general expressions maybe simplified for three different regions of $T$:

I. $\mu > m,\ T \ll \mu -m$
\begin{eqnarray}
 \Delta\overline{\sigma}_i &=& \int_m^{\mu} f_i (z) dz + \frac{\pi^2}{6} T^2 f'_i(\mu)\ann
 &+& \frac{7\pi^4}{360} T^4 f'''_i (\mu) + O(e^{-\frac{\mu-m}{T}}).
\end{eqnarray}

II. $\mu < m,\ T \ll m - \mu$
\begin{equation}
 \Delta\overline{\sigma}_i = O(e^{-\frac{m -\mu}{T}}). \label{eq:apBnonzero}
\end{equation}

III. $\mu = m,\ T \ll m$

\begin{eqnarray}
 \Delta\overline{\sigma}_i &=& T \ln 2 f_i(m) + \frac{\pi^2}{12} T^2 f'_i(m) + \frac{3}{4}\zeta_R (3) T^3 f''_i(m) \ann
 &+&\frac{7\pi^4}{720} T^4 f'''_i(m) + O(e^{-\frac{m}{T}}).
\end{eqnarray}

In gapeless case, $m=0$, 
\begin{eqnarray}
 	f_\te (z) &=& \frac{8}{\pi \lambda} \Re \frac{q\left(q^2  - \lambda^2 \right)}{r_0 (qk_F + \lambda r_0)} ,\ann 
 	f_\tm (z) &=& \frac{8}{\pi k_F} \Re \frac{q r_0 - \lambda k_F}{k_F r_0 + q \lambda},
\end{eqnarray}
with $r_0 = \sqrt{q^2 + k^2 v_F^2}, q=\lambda-2\ii z$. We have two regions

I. $\mu > 0,\ T \ll \mu$
\begin{eqnarray}
 \Delta\overline{\sigma}_i &=& \int_0^{\mu} f_i (z) dz + \frac{\pi^2}{6} T^2 f'_i(\mu)\ann
 &+& \frac{7\pi^4}{360} T^4 f'''_i (\mu) + O(e^{-\frac{\mu}{T}}).
\end{eqnarray}

II. $\mu = 0,\ T\to 0$

\begin{equation}
 \Delta\overline{\sigma}_i =  \frac{3}{2}\zeta_R (3) T^3 f''_i(0) + \frac{15}{8} \zeta_R (5) T^5 f^{(4)}_i(0) + \ldots . \label{eq:apBzeros}
\end{equation}
\input{cond.bbl.tex}

%\bibliography{/home/nail/MEGAsync/Bib/my_publ,/home/nail/MEGAsync/Bib/books,/home/nail/MEGAsync/Bib/gravity,%
%/home/nail/MEGAsync/Bib/casimir,/home/nail/MEGAsync/Bib/topiso}
%%\bibliography{ref}
%\bibliographystyle{apsrev4-1}
\end{document}